\documentclass[reprint]{revtex4-2}

\usepackage{xcolor}
\usepackage[british]{babel}
\usepackage[utf8]{inputenc}
\usepackage{epsfig}
\usepackage{graphicx}
\usepackage{physics}
\usepackage{enumitem}

\usepackage{subfigure}
\usepackage{tikz}
\usetikzlibrary{shapes,arrows,arrows.meta,positioning,calc}
\usepackage{pgf}
\usepackage{ifpdf}

\ifpdf
  \usetikzlibrary{arrows.meta}
\else
  \PassOptionsToPackage{dvips}{graphicx}
\fi
\tikzset{
    line/.style = {
        draw,
        -{Latex[round,length=20pt,width=6pt]}
    },
    cloud/.style = {
        draw, ellipse, node distance=2.5cm,
        minimum height=2em
    }
}

\usepackage{amsmath,amssymb,amsthm,mathtools}\usepackage{bm}
\usepackage{geometry}
\usepackage{microtype}
\usepackage[colorlinks=true,linkcolor=blue,citecolor=blue,urlcolor=blue]{hyperref}
\theoremstyle{remark}

\begin{document}

\title{Can Quantum Field Theory be Recovered from Time-Symmetric Stochastic Mechanics? Part II: Prospects for a Trajectory Interpretation}
\author{Simon Friederich, Mritunjay Tyagi}
\email{s.m.friederich@rug.nl, m.tyagi@rug.nl}
\affiliation{University of Groningen, University College Groningen, Hoendiepskade 23/24, 9718BG Groningen, the Netherlands}

\begin{abstract}
In a companion paper we derived a unique time-reversal-invariant stochastic generalization of the Liouville equation and showed that it coincides with the evolution equation for the Husimi $Q$-function in a broad class of bosonic quantum field theories. Here we investigate the prospects for interpreting that evolution equation in terms of underlying stochastic trajectories. Drawing on Drummond's time-symmetric stochastic action formalism, we show that the traceless diffusion Fokker-Planck equation defines a natural measure over stochastic trajectories conditional on mixed-time boundary conditions. However, we identify a significant gap: it has not been established that every $Q$-function can be represented as a weighted average of these conditional probabilities over boundary values. The trajectory interpretation holds for ensembles with fixed boundary conditions but does not straightforwardly extend to arbitrary quantum states. Despite this limitation, we show that Drummond's trajectory dynamics are fundamentally non-Markovian -- a natural consequence of combining stochasticity with time-reversal invariance. This non-Markovianity places the dynamics outside the scope of the ontological models framework and thereby explains why the major no-go theorems for hidden-variable theories do not rule out the approach. These results clarify both the achievements and the remaining challenges in the project of understanding quantum field theory as the statistical mechanics of time-symmetric stochastic processes.
\end{abstract}

\maketitle

\section{Introduction}\label{sec:intro}

In a companion paper \citep{friederichtyagi_PartI}, we motivated a stochastic generalization of the classical Liouville equation by imposing physically motivated constraints: reduction to classical dynamics as $\hbar\to 0$, Fokker-Planck form, local Hamiltonian dependence, time-reversal invariance, energy conservation, and minimality. The resulting evolution equation features a traceless diffusion matrix and coincides exactly with the Schr\"odinger equation in the coherent-state representation -- i.e.\ the evolution equation for the Husimi $Q$-function -- for a broad class of bosonic quantum field theories whose Anti-Wick Hamiltonian symbols are at most quadratic in each complex phase-space variable.

This coincidence suggests interpreting the Husimi function as a genuine probability density on phase space, with all dynamical variables having sharp values at all times. If such an interpretation can be sustained, it would solve the quantum measurement problem: ``collapse'' would be nothing more than Bayesian updating upon learning a measurement outcome. However, realizing this vision requires establishing that the evolution equation admits an interpretation in terms of underlying stochastic trajectories.

In this paper, we investigate the prospects for such a trajectory interpretation, with a focus on the achievements and limitations of the approach. After reviewing the Husimi function and the evolution equation it satisfies (Section~\ref{sec:husimi_and_evolution}), we develop the framework for a trajectory interpretation in Section~\ref{sec:trajectories}. We first review the standard case of positive-definite diffusion, introducing a hierarchy of ensembles ($E_0$--$E_3$) that clarifies what a trajectory interpretation must achieve. We then review and assess Drummond's construction \citep{drummond2021}, which provides the trajectory interpretation at the level of ensembles with fixed boundary conditions ($E_2$), and identify the ``representability gap'' -- the fact that it has not been established that every $Q$-function can be written as such an average over boundary conditions.

In Sections~\ref{sec:non-Markovian} and \ref{sec:no_go}, we show that the resulting trajectory dynamics are fundamentally non-Markovian, which places them outside the ontological models framework and renders the major no-go theorems (Bell, PBR, Spekkens contextuality and negativity) inapplicable. This is an important positive result: the real obstacles to the project are the representability gap and the extension to all Standard Model Hamiltonians, not the no-go theorems.

We conclude in Section~\ref{sec:summary} with a discussion of possible routes forward.

\section{The Husimi function and its evolution}\label{sec:husimi_and_evolution}

\subsection{The Husimi function as a probability density}

The Husimi $Q$-function \citep{husimi1940} is a phase-space representation of the quantum density matrix $\hat{\rho}$, defined by
\begin{equation}
Q(\boldsymbol{\alpha}, \boldsymbol{\alpha}^*,t) = \frac{1}{\pi^N}\bra{\boldsymbol{\alpha}}\hat{\rho}(\hat{a}, \hat{a}^\dagger,t)\ket{\boldsymbol{\alpha}}\,,
\label{husimi}
\end{equation}
where $\ket{\boldsymbol{\alpha}}$ are coherent states labeled by complex phase-space coordinates $\boldsymbol{\alpha}$.

Among the commonly used phase-space representations of the quantum state -- alongside the Wigner function and the Glauber-Sudarshan $P$-function -- the Husimi function is uniquely predestined to play the role of a probability density on phase space. It is everywhere non-negative and normalized to unity: $\int Q(\boldsymbol{\alpha})\,d\boldsymbol{\alpha} = 1$. Moreover, it provides quantum expectation values via the simple phase-space integral
\begin{equation}
\mathrm{Tr}(\hat{A}\hat{\rho}) = \int A_{\mathrm{aW}}(\boldsymbol{\alpha})\,Q(\boldsymbol{\alpha})\,d\boldsymbol{\alpha}\,,
\label{expectation}
\end{equation}
provided $A_{\mathrm{aW}}(\boldsymbol{\alpha})$ is the Anti-Wick symbol of the operator $\hat{A}$ \citep{lee,folland1989}. This means that the Husimi function, when interpreted as a probability density, gives the correct quantum expectation values for \emph{all} observables, provided one uses Anti-Wick quantization to map between phase-space functions and operators.

The idea to interpret the Husimi function as a genuine probability density has a long history. It appears to have been proposed first by Bopp \citep{bopp1956} and has been revived from physical \citep{drummondreid2020} and philosophical \citep{friederich2024} perspectives. Intriguingly, this move allows one to sidestep the Kochen-Specker theorem \citep{kochen1967} as an obstacle towards assigning sharp values to all observables. Importantly, the Husimi function provides the correct measurement statistics in heterodyne detection, which simultaneously probes conjugate quadratures \citep{PhysRevLett.117.070801,Wiseman_Milburn_2009}. Appleby has shown that it provides the distribution of results in ``retrodictively optimal'' phase-space measurements \citep{appleby1998optimaljointmeasurementsposition}. It also underlies the phase-space formulations of Bohmian mechanics that treat position and momentum democratically \citep{depolavieja1996}. These observations suggest -- though admittedly do not conclusively establish -- that an interpretation of the Husimi function as a genuine probability density on phase space may be empirically adequate.

\subsection{The general evolution equation}

The time evolution of $Q$ is the Schr\"odinger (or von Neumann) equation in the coherent-state representation. In terms of the Anti-Wick Hamiltonian symbol $H_{\mathrm{aW}}$, the general evolution equation is \citep{tyagifriederich}:
\begin{equation}
\partial_t Q
= \frac{i}{\hbar}\sum_{|\mathbf{m}|\ge1}\frac{\hbar^{|\mathbf{m}|}}{\mathbf{m}!}
\Big[
\partial^{\mathbf{m}}\!\bigl(\bar{\partial}^{\mathbf{m}} H_{\mathrm{aW}}\cdot Q\bigr)
-
\bar{\partial}^{\mathbf{m}}\!\bigl(\partial^{\mathbf{m}} H_{\mathrm{aW}}\cdot Q\bigr)
\Big]\,,\label{full_Q_evolution}
\end{equation}
where $\mathbf{m}$ is a multi-index, $\partial^{\mathbf{m}} = \prod_i (\partial/\partial\alpha_i)^{m_i}$, and $\bar{\partial}^{\mathbf{m}} = \prod_i (\partial/\partial\alpha_i^*)^{m_i}$.

This infinite series \emph{truncates} when the Anti-Wick Hamiltonian symbol has limited polynomial degree. If $H_{\mathrm{aW}}$ is at most quadratic in each complex variable $\alpha_i$ (and its conjugate), the series truncates at second order, yielding the Fokker-Planck form:
\begin{widetext}
\begin{align}
\frac{\partial Q}{\partial t}
&= i\sum_{i}
\left[
\frac{\partial}{\partial \alpha_i}
\Bigl(\frac{\partial H_{\mathrm{aW}}}{\partial \alpha_i^*}\,Q\Bigr)
- \frac{\partial}{\partial \alpha_i^*}
\Bigl(\frac{\partial H_{\mathrm{aW}}}{\partial \alpha_i}\,Q\Bigr)
\right]
+ \frac{i\hbar}{2} \sum_{i,j}
\left\{
\frac{\partial^2}{\partial \alpha_i \partial \alpha_j}
\Bigl(\frac{\partial^2 H_{\mathrm{aW}}}{\partial \alpha_i^*\partial \alpha_j^*}\,Q\Bigr)
- \frac{\partial^2}{\partial \alpha_i^* \partial \alpha_j^*}
\Bigl(\frac{\partial^2 H_{\mathrm{aW}}}{\partial \alpha_i\partial \alpha_j}\,Q\Bigr)
\right\}\!.
\label{Q_evolution_quadratic}
\end{align}
\end{widetext}
This is the traceless diffusion Fokker-Planck equation (FPE). In dimensionless real phase space coordinates, the second term (the ``diffusion'' contribution) takes the form
\begin{equation}
\bigl(\partial_t Q\bigr)_{\mathrm{diff}} = \frac{1}{4\hbar}\sum_{a,b=1}^{2N}\tilde{\partial}_a \tilde{\partial}_b\Bigl(\bigl[\tilde{H}'',\,J\bigr]_{ab}\,Q\Bigr)\,.\label{diffusion_components}
\end{equation}
where $[\tilde{H}'',J]$ is the traceless diffusion matrix and $\tilde{\partial}_a,\, \tilde{\partial}_b$ are derivatives with respect to dimensionless phase space coordinates (see \citep{friederichtyagi_PartI} for details).

As shown in \citep{friederichtyagi_PartI}, this equation can be motivated based on physical constraints such as time reversal invariance, energy conservation, minimality, and reduction to the Liouville equation in the classical limit -- a striking coincidence that may be taken to further suggest interpreting $Q$ as a genuine probability density on phase space.

\subsection{Drummond's trajectory proposal}

The traceless character of the diffusion matrix -- with equal numbers of positive and negative eigenvalues -- precludes the standard forward-time trajectory interpretation, which requires positive definite diffusion. However, Drummond \citep{drummond2021} has proposed a trajectory interpretation based on \emph{mixed-time boundary conditions}, exploiting the fact that the positive and negative diffusion directions can be associated with forward and backward propagation in time, respectively. We now investigate this proposal in detail.

\section{Trajectory interpretation}\label{sec:trajectories}

\subsection{The standard case: positive definite diffusion}
\label{subsec:standard}

To establish what a trajectory interpretation of the traceless diffusion FPE needs to achieve, it is instructive to first review the well-understood case of an ordinary Fokker-Planck equation with positive definite diffusion:
\begin{eqnarray}
\frac{\partial \rho(\mathbf{x},t)}{\partial t}&=&-\sum_i\frac{\partial }
{\partial x_i}\bigl[a_i(\mathbf{x}) \rho\bigr] \nonumber\\&+& 
\sum_{i,j}\frac{1}{2} \frac{\partial^2}{\partial x_i\partial x_j}
\bigl[D_{ij}(\mathbf{x}) \rho\bigr]\,,
\label{FP_recap}
\end{eqnarray}
where $D_{ij}$ is positive definite. When $D_{ij}$ is positive definite, this equation determines a measure over future trajectories 
starting from an arbitrary initial configuration $\mathbf{x}_0$.

Specifically, these paths satisfy the stochastic differential equation:
\begin{equation}
dx_i(t) = a_i(\mathbf{x}(t))\,dt + c_{ij}(\mathbf{x}(t))\,dW_j(t)\,,
\label{SDE_recap}
\end{equation}
where $D_{ij} = \sum_k c_{ik}c_{jk}$ and $dW_j$ are increments of 
independent Wiener processes. Given an initial condition 
$\mathbf{x}(t_0) = \mathbf{x}_0$, Eq.~(\ref{SDE_recap}) defines a 
stochastic process with random trajectories $\mathbf{x}(t)$ for 
$t > t_0$.

The probability distribution over these trajectories can be characterized 
using the \emph{Onsager-Machlup functional} 
\citep{onsagermachlup1953}. For a smooth path $\mathbf{x}(t)$, this 
functional takes the form
\begin{equation}
S[\mathbf{x}] = \int_{t_0}^{t_f} \mathcal{L}_{\mathrm{OM}}
(\mathbf{x}, \dot{\mathbf{x}})\,dt\,,
\end{equation}
where the Onsager-Machlup Lagrangian is
\begin{eqnarray}
&&\mathcal{L}_{\mathrm{OM}}(\mathbf{x}, \dot{\mathbf{x}}) 
\label{OM_lagrangian}\\
&&\quad= \sum_{i,j}\frac{1}{2}D^{-1}_{ij}(\mathbf{x})
(\dot{x}_i - a_i(\mathbf{x}))(\dot{x}_j - a_j(\mathbf{x})) + 
V(\mathbf{x})\,,\nonumber
\end{eqnarray}
with $D^{-1}_{ij}$ the inverse of the diffusion matrix. The potential term 
$V(\mathbf{x})$ depends on both the drift and the spatial variation of the 
diffusion; for constant diffusion it reduces to 
$V(\mathbf{x}) = -\frac{1}{2}\sum_i \partial_i a^i(\mathbf{x})$, 
but in general involves additional geometric terms from 
$\partial_k D_{ij}(\mathbf{x})$ \citep{graham1977}.
The Onsager-Machlup Lagrangian has a transparent physical meaning: the 
kinetic-like term $\frac{1}{2}D^{-1}_{ij}(\dot{x}_i - a_i)
(\dot{x}_j - a_j)$ penalizes deviations of the trajectory's velocity from 
the deterministic drift $\mathbf{a}(\mathbf{x})$. Trajectories that 
closely follow the drift are exponentially more probable than those that 
deviate strongly. Technically, smooth paths have measure zero in the space 
of all continuous paths; the Onsager-Machlup functional should be 
understood as defining the limiting probability density for paths within a 
tube of radius $\epsilon$ around the smooth path, in the limit 
$\epsilon \to 0$.

Using the Onsager-Machlup functional, the conditional probability for the 
system to reach $\mathbf{x}_f$ at time $t_f$ given that it started at 
$\mathbf{x}_0$ at time $t_0$ can be expressed as a path integral:
\begin{equation}
P(\mathbf{x}_f, t_f | \mathbf{x}_0, t_0) = 
\int_{\mathbf{x}(t_0)=\mathbf{x}_0}^{\mathbf{x}(t_f)=\mathbf{x}_f} 
\mathcal{D}[\mathbf{x}]\,e^{-S[\mathbf{x}]}\,,
\label{forward_conditional}
\end{equation}
where the path integral is over all continuous trajectories connecting 
$\mathbf{x}_0$ at $t_0$ to $\mathbf{x}_f$ at $t_f$.

We now frame this trajectory interpretation in terms of ensembles, which 
will facilitate the generalization to traceless diffusion.

\textbf{Ensemble $E_1(\mathbf{x}_0)$.} The ensemble of all trajectories 
emanating from $\mathbf{x}_0$, weighted by the Onsager-Machlup measure. 
The probability density $\rho_{E_1}(\mathbf{x},t)$ computed from 
$E_1(\mathbf{x}_0)$ satisfies the FPE with initial condition 
$\delta(\mathbf{x} - \mathbf{x}_0)$.

\textbf{Ensemble $E_2$.} Average over initial conditions with a 
distribution $P(\mathbf{x}_0)$:
\begin{equation}
\rho_{E_2}(\mathbf{x},t) = \int P(\mathbf{x}, t | \mathbf{x}_0, t_0)\, 
P(\mathbf{x}_0)\, d\mathbf{x}_0\,.
\label{eq:E2_standard}
\end{equation}
Since the FPE is linear, $\rho_{E_2}$ also satisfies it, with initial 
condition $\rho_{E_2}(\mathbf{x},t_0) = P(\mathbf{x}_0)$.

The trajectory interpretation for the standard FPE is \emph{complete} 
because \emph{every} solution $\rho$ can be expressed in the form 
Eq.~(\ref{eq:E2_standard}): simply take 
$P(\mathbf{x}_0) = \rho(\mathbf{x}_0, t_0)$. In the language we develop below, the $E_2$ ensemble already provides a trajectory interpretation for any target probability density -- no further conditioning ($E_3$) is needed, because every initial condition is representable. This will turn out to be the crucial point of contrast with the traceless diffusion case.

\subsection{The general strategy for traceless diffusion}\label{subsec:general_strategy}

For the traceless diffusion FPE, the standard construction fails: the diffusion matrix has both positive and negative eigenvalues, and the forward-time initial value problem is ill-posed \citep{miranker1961}. However, the hierarchy of ensembles suggests a general strategy.

Suppose we can define a measure over trajectories conditional on boundary conditions that involve both an initial and a final time -- a natural expectation when the diffusion has both forward and backward components. Let $\boldsymbol{\phi}_{\mathrm{IN}}$ denote these boundary conditions (whose precise form will be specified below). Then we can attempt to construct:
\begin{itemize}
\item \textbf{Ensemble $E_0$}: All paths (within some suitably chosen reference class) consistent with the boundary specification $\boldsymbol{\phi}_{\mathrm{IN}}$.
\item \textbf{Ensemble $E_1$}: A subensemble weighted by an appropriate path measure. One may try to identify a path measure here for which the synchronous probability density satisfies the traceless diffusion FPE with appropriate (mixed-time) boundary conditions $\boldsymbol{\phi}_{\mathrm{IN}}$.
\item \textbf{Ensemble $E_2$}: Obtained by averaging $E_1$ ensembles over boundary conditions with a distribution $P(\boldsymbol{\phi}_{\mathrm{IN}})$:
\begin{equation}
\rho(\boldsymbol{\phi},t) = \int G(\boldsymbol{\phi},t|\boldsymbol{\phi}_{\mathrm{IN}})\,P(\boldsymbol{\phi}_{\mathrm{IN}})\,d\boldsymbol{\phi}_{\mathrm{IN}}\,.
\label{eq:E2_general}
\end{equation}
If the $E_1$-ensemble satisfies the traceless diffusion FPE, this $\rho$ will automatically satisfy the FPE.
\end{itemize}

The trajectory interpretation succeeds if any Husimi function can be written in the form Eq.~(\ref{eq:E2_general}) for some choice of boundary distribution. In this case, the Husimi function evolution can be seen as reflecting the time dependence of the measure over underlying paths in the ensemble $E_2$.

Note that this general strategy does not presuppose the specific form of the boundary conditions $\boldsymbol{\phi}_{\mathrm{IN}}$, nor the specific coordinates in which the path measure is defined. A Schr\"odinger bridge formulation \citep{schrodinger1931,leonard2014}, with the entire phase space configuration fixed at both temporal ends, would be another option. What matters is that the path measure leads to a propagator $G$ whose weighted averages satisfy the FPE and that the Husimi function can be represented in the form Eq.~(\ref{eq:E2_general}).

We now turn to Drummond's specific realization of this strategy.

\subsection{Drummond's construction}\label{subsec:drummond}

Drummond's key insight \citep{drummond2021} is to introduce phase-space coordinates $\boldsymbol{\phi} = (\mathbf{x}, \mathbf{y})$ in which the traceless diffusion matrix takes a particularly simple form: diagonal and constant (i.e. no dependence on phase space location), with the $\mathbf{x}$ components having positive diffusion and the $\mathbf{y}$ components having negative diffusion:
\begin{equation}
D = \begin{pmatrix} d^x & 0 \\ 0 & -d^y \end{pmatrix}\,,
\label{eq:D_diagonal}
\end{equation}
where $d^x$ and $d^y$ are positive-definite matrices whose only eigenvalue is $d$, the constant value of diffusion.

The transformation from the original complex mode amplitudes $\boldsymbol{\alpha}$ to the variables $\boldsymbol{\phi} = (\mathbf{x}, \mathbf{y})$ proceeds in two steps \citep[Sect.~II.F and III]{drummond2021}. First, a (generally nonlinear) change of variables is performed to make the diffusion constant in phase space, absorbing any field-dependence. This is achieved by a logarithmic transformation. Second, a linear mapping to real quadrature variables is performed to diagonalize the diffusion into positive and negative definite blocks. For general Hamiltonians, where the diffusion depends on phase-space location, the first step may not yield constant diffusion; the path integral formalism still applies \citep{graham1977}, but the resulting expressions are more involved. We refer to \citep{drummond2021} for the details and work with the general form Eq.~(\ref{eq:D_diagonal}) in what follows.

In these coordinates, the time-symmetric stochastic differential equation (TSSDE) takes the form:
\begin{align}
\mathbf{x}(t) &= \mathbf{x}_0 + \int_{t_0}^{t} \mathbf{a}^x(\boldsymbol{\phi}(t'))\,dt' + \int_{t_0}^{t} d\mathbf{w}^x\,,\nonumber\\
\mathbf{y}(t) &= \mathbf{y}_f + \int_{t_f}^{t} \mathbf{a}^y(\boldsymbol{\phi}(t'))\,dt' + \int_{t_f}^{t} d\mathbf{w}^y\,,
\label{eq:TSSDE}
\end{align}
where the two sets of coordinates propagate in opposite time directions, and the independent Gaussian noise terms satisfy $\langle dw^x_i dw^x_j \rangle = d^x_{ij}\,dt$, $\langle dw^y_i dw^y_j \rangle = d^y_{ij}\,dt$, $\langle dw^x_i dw^y_j \rangle = 0$.

The boundary conditions are of mixed-time type: $\mathbf{x}(t_0) = \mathbf{x}_0$ is fixed at the initial time, while $\mathbf{y}(t_f) = \mathbf{y}_f$ is fixed at the final time. The ``input'' is $\boldsymbol{\phi}_{\mathrm{IN}} = (\mathbf{x}_0, \mathbf{y}_f)$.

The path probability for a trajectory $\boldsymbol{\phi}(t)$ connecting 
the boundary values is determined by a real action functional 
$S[\boldsymbol{\phi}] = \int_{t_0}^{t_f} L(\boldsymbol{\phi}, 
\dot{\boldsymbol{\phi}})\,dt$. The Lagrangian is of 
\emph{Onsager-Machlup} type \citep{onsagermachlup1953,graham1977} -- the 
same functional form that characterizes path probabilities in standard 
forward-time diffusion processes -- but now applied to a time-symmetric 
setting with mixed-time boundary conditions. Explicitly 
\citep[Eq.~(111)]{drummond2021}:
\begin{equation}
L = \sum_{\mu} \frac{1}{2d}(\dot{\phi}_{\mu} - A_{\mu}(\boldsymbol{\phi}))^2 
- V(\boldsymbol{\phi})\,,
\label{eq:symmetric_OM}
\end{equation}
with $V(\boldsymbol{\phi}) = -\frac{1}{2}\sum_{\mu}\partial_{\mu} 
A^{\mu}(\boldsymbol{\phi})$, where $d$ is the constant diffusion magnitude 
and $A_{\mu}$ are the drift components in the $\boldsymbol{\phi}$ 
coordinates. The first term penalizes deviations of the trajectory's 
velocity from the drift -- trajectories that follow the drift closely are 
exponentially more probable than those that deviate -- while $V$ is a 
potential term arising from the divergence of the drift. The crucial 
property is that $S$ is \emph{real}, not imaginary as in the Feynman path 
integral. Provided the corresponding path integral is normalizable this allows the expression
\begin{equation}
\mathbb{P}[\boldsymbol{\phi}(\cdot) | \mathbf{x}_0, \mathbf{y}_f] \propto 
\exp\left(-S[\boldsymbol{\phi}]\right)
\label{eq:path_measure}
\end{equation}
to define a proper (normalizable) probability measure over trajectories, with the most probable trajectories being those that minimize the action.

Drummond does not use the term ``Onsager-Machlup,'' but the formal 
structure is precisely that of an Onsager-Machlup functional for the 
time-symmetric stochastic process defined by Eq.~(\ref{eq:TSSDE}).

Drummond defines a time-symmetric propagator (TSP) as the solution to the FPE with delta-function boundary conditions at both temporal ends:
\begin{align}
G_x(\mathbf{x},t_0|\boldsymbol{\phi}_{\mathrm{IN}}) &= \delta(\mathbf{x} - \mathbf{x}_0)\,,\nonumber\\
G_y(\mathbf{y},t_f|\boldsymbol{\phi}_{\mathrm{IN}}) &= \delta(\mathbf{y} - \mathbf{y}_f)\,.
\label{eq:TSP_boundary}
\end{align}

In the language of Section~\ref{subsec:general_strategy}, this construction provides the $E_1$ ensembles (trajectories weighted by the path measure for fixed boundary conditions $\boldsymbol{\phi}_{\mathrm{IN}}$) and the $E_2$ ensembles (averages over boundary conditions). Any $\rho$ of the form
\begin{equation}
\rho(\boldsymbol{\phi},t) = \int G(\boldsymbol{\phi},t|\boldsymbol{\phi}_{\mathrm{IN}})\,P(\boldsymbol{\phi}_{\mathrm{IN}})\,d\boldsymbol{\phi}_{\mathrm{IN}}
\label{eq:rho_from_G}
\end{equation}
satisfies the FPE. This may be seen as Drummond's central achievement in \citep{drummond2021}.

\subsection{Factorization over intermediate times}\label{subsec:factorization}

An important structural property of the conditional probabilities from Eq.~(\ref{eq:path_measure}) is that they factorize over intermediate times. Using the additivity of the action, $S[\boldsymbol{\phi}]_{t_i}^{t_f} = S[\boldsymbol{\phi}]_{t_i}^{t} + S[\boldsymbol{\phi}]_{t}^{t_f}$, one obtains:
\begin{align}
&P(\mathbf{x}_f, \mathbf{x}, \mathbf{y}, \mathbf{y}_i|\mathbf{y}_f, \mathbf{x}_i)
\label{eq:bridge_symm}\\
&\qquad\propto P(\mathbf{x}, \mathbf{y}_i|\mathbf{y}, \mathbf{x}_i) \cdot P(\mathbf{x}_f, \mathbf{y}|\mathbf{y}_f, \mathbf{x})\,,\nonumber
\end{align}
where the proportionality involves a normalization factor that may depend on the boundary values $\mathbf{x}_i$ and $\mathbf{y}_f$. While reminiscent of the Chapman-Kolmogorov equation for Markov processes, this factorization actually corresponds to non-Markovian dynamics, as we show in Section~\ref{sec:non-Markovian}.

\subsection{The representability gap}\label{subsec:representability_gap}

Given that any $\rho$ of the form Eq.~(\ref{eq:rho_from_G}) satisfies the FPE, one might expect that every $Q$-function can be expressed in this form. If so, the trajectory interpretation would extend to all quantum states. However, this expectation is not generally warranted, as Drummond himself notes \citep[Sect.~III.B]{drummond2021}. We have:
\begin{enumerate}[label=(\roman*)]
\item If $\rho$ is of the form Eq.~(\ref{eq:rho_from_G}), then $\rho$ satisfies the FPE.
\item Every $Q$-function satisfies the FPE.
\end{enumerate}
It does \emph{not} follow that every $Q$-function can be written in the form Eq.~(\ref{eq:rho_from_G}). Inferring this would be a case of the fallacy sometimes called \emph{affirming the consequent}.

One might attempt to resolve this by starting with some $E_2$ ensemble and conditioning on an initial $Q$-function -- accepting or rejecting trajectories based on their initial phase-space location. This yields an ensemble $E_3$ that does have a trajectory interpretation (being a subensemble of trajectories). However, the probability density for $E_3$ will in general \emph{not} satisfy the FPE. The overall ensemble $E_2$ satisfies the FPE because of how it was constructed -- through the specific Onsager-Machlup weighting and integration over boundary values. Conditioning on initial configurations disrupts this balance.

One might protest: ``But we \emph{know} that $Q$ satisfies the FPE!'' Indeed we know this: the Husimi function satisfies the FPE by construction, being equivalent to the Schr\"odinger equation. However, the problem is that this equation and the time evolution of the probability density of an $E_3$ ensemble are in general \emph{different} evolutions, which happen to share an initial condition. The ensemble $E_3$ provides a trajectory interpretation for something -- but that something may not be the quantum field theory with a Husimi function that evolves according to the Schr\"odinger equation.

We summarize:
\begin{table*}[t]
\centering
\begin{tabular}{p{7cm}cc}
\hline
& Trajectory interpretation & Satisfies FPE \\
\hline
$E_0$ (generic paths, fixed boundary conditions) & Yes & In general, No \\
$E_1$ (OM-weighted, fixed boundary conditions) & Yes & Yes \\
$E_2$ (averaged over boundary conditions) & Yes & Yes \\
$E_3$ (conditioned on initial $Q$) & Yes & In general, No \\
$Q$-evolution & would require $E_3$ matching & Yes \\
\hline
\end{tabular}
\caption{Hierarchy of ensembles and the status of the trajectory interpretation. The FPE is satisfied at the $E_1$ and $E_2$ levels, but this is not where it is needed: the trajectory interpretation for quantum field theory requires that the $Q$-function evolution be recovered at the $E_3$ level, where an initial $Q$-function has been imposed by conditioning. This has not been established.}
\label{tab:ensembles}
\end{table*}

In order for a trajectory interpretation for quantum field theory to be established along the lines of Drummond's construction, one would have to show that any Husimi-function admits a representation via some $P(\boldsymbol{\phi}_{\mathrm{IN}})$. This would mean that the $E_3$ level reduces to the $E_2$ level. Unfortunately, we see no robust reason to expect that this scenario is the case in general.

An important remark is in order. Unlike (perhaps) the situation for the Standard Model coverage discussed in the companion paper \citep{friederichtyagi_PartI}, the representability gap is \emph{not} a limitation that could be circumvented by reformulating the theory. The Husimi function evolution must be recoverable at the $E_3$ level, not merely at the $E_2$ level, for a trajectory interpretation to be viable. This makes closing the representability gap -- by proving that every $Q$-function admits a representation in the form Eq.~(\ref{eq:rho_from_G}) -- the central mathematical challenge that would have to be addressed in order to provide a trajectory interpretation for quantum theories with time evolution corresponding to the traceless diffusion FPE along the lines of Drummond's construction in \citep{drummond2021}.

In certain special cases, such as the parametric amplification Hamiltonian where the $\mathbf{x}$ and $\mathbf{y}$ coordinates evolve independently \citep{drummond2021,drummondreid2020}, the problem simplifies considerably and there may be no representability gap. For the general case with coupled dynamics, however, no proof is available, and Drummond himself appears skeptical about its solvability in general \citep[Sect.~III.B]{drummond2021}.

\subsection{Significance}

Despite the representability gap, Drummond's construction has substantial value. It demonstrates that the traceless diffusion FPE is \emph{compatible} with a trajectory picture -- a non-trivial fact given that the standard forward-time construction completely fails. The path measure is well-defined for any set of mixed-time boundary conditions, the action is real, and the resulting propagator satisfies the FPE. Furthermore, the structural features of the trajectory dynamics -- particularly their non-Markovian character, to which we now turn -- hold regardless of whether the representability gap can be closed, and have important implications for the relation to no-go theorems.

\section{Non-Markovian dynamics}\label{sec:non-Markovian}

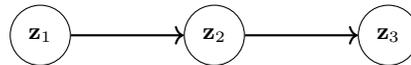
\begin{figure}[htbp]
\centering
\begin{tikzpicture}[node distance = 1.5cm, auto]
\node [circle, draw, minimum size=0.8cm] (z1) {$\mathbf{z}_1$};
\node [circle, draw, minimum size=0.8cm, right= of z1] (z2) {$\mathbf{z}_2$};
\node [circle, draw, minimum size=0.8cm, right= of z2] (z3) {$\mathbf{z}_3$};
\draw [->, thick] (z1) -- (z2);
\draw [->, thick] (z2) -- (z3);
\end{tikzpicture}
\caption{Dependency structure for a Markov process.}
\label{fig:Markov}
\end{figure}

A striking feature of the time-symmetric stochastic dynamics is the failure of the Markov condition. In a Markov process, the multi-step transition probability factorizes as
\begin{equation}
P(\mathbf{z}_3, \mathbf{z}_2|\mathbf{z}_1) = P(\mathbf{z}_3|\mathbf{z}_2)\,P(\mathbf{z}_2|\mathbf{z}_1)\,,
\label{eq:Markov_factor}
\end{equation}
so that the intermediate configuration $\mathbf{z}_2$ ``screens off'' all correlations between past and future (Fig.~\ref{fig:Markov}).

\begin{figure}[htbp]
\centering
\begin{tikzpicture}[node distance = 1.5cm, auto]
\node [circle, draw, minimum size=0.8cm] (x1) {$\mathbf{x}_1$};
\node [circle, draw, minimum size=0.8cm, right= of x1] (x2) {$\mathbf{x}_2$};
\node [circle, draw, minimum size=0.8cm, right= of x2] (x3) {$\mathbf{x}_3$};
\node [circle, draw, minimum size=0.8cm, below of = x1] (y1) {$\mathbf{y}_1$};
\node [circle, draw, minimum size=0.8cm, right= of y1] (y2) {$\mathbf{y}_2$};
\node [circle, draw, minimum size=0.8cm, right= of y2] (y3) {$\mathbf{y}_3$};

\draw [->, thick] (x1) -- (x2);
\draw [->, thick] (x2) -- (x3);
\draw [->, thick] (x1) -- (y1);
\draw [->, thick] (y3) -- (x3);
\draw [->, thick] (y2) -- (y1);
\draw [->, thick] (y3) -- (y2);
\draw [->, thick] (x2) to [bend right=35] (y2);
\draw [->, thick] (y2) to [bend right=35] (x2);
\end{tikzpicture}
\caption{Intuitively visualized dependency structure for time-symmetric dynamics (not a rigorous causal graph). The cyclic arrows between $\mathbf{x}_2$ and $\mathbf{y}_2$ violate the acyclicity required for Markov processes.}
\label{fig:non_Markov}
\end{figure}
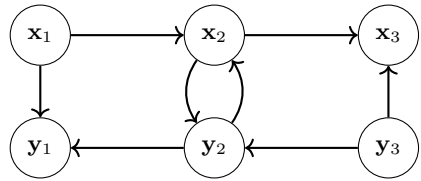

In contrast, the time-symmetric dynamics factorizes as in Eq.~(\ref{eq:bridge_symm}):
\begin{align}
&P(\mathbf{x}_3, \mathbf{x}_2, \mathbf{y}_2, \mathbf{y}_1|\mathbf{y}_3, \mathbf{x}_1)
\label{eq:non_Markov}\\
&\quad\propto P(\mathbf{x}_2, \mathbf{y}_1|\mathbf{y}_2, \mathbf{x}_1) \cdot P(\mathbf{x}_3, \mathbf{y}_2|\mathbf{y}_3, \mathbf{x}_2)\,.\nonumber
\end{align}
The crucial difference is that $\mathbf{y}_2$ appears on opposite sides of the conditional stroke in the two factors: it is an output in the first factor but an input in the second, and vice versa for $\mathbf{x}_2$. This creates a cyclic dependency structure (intuitively visualized in Fig.~\ref{fig:non_Markov}) that cannot be represented by a directed acyclic graph.

\subsection{Failure of conditional independence}
 
To demonstrate that the Markov screening-off property fails, we examine whether conditioning on $\boldsymbol{\phi}_2 = (\mathbf{x}_2, \mathbf{y}_2)$ renders $\mathbf{x}_1$ and $\mathbf{y}_3$ independent, i.e. whether
\begin{equation}
P(\mathbf{x}_1, \mathbf{y}_3 \mid \mathbf{x}_2, \mathbf{y}_2) = P(\mathbf{x}_1 \mid \mathbf{x}_2, \mathbf{y}_2)\times P( \mathbf{y}_3 \mid \mathbf{x}_2, \mathbf{y}_2)\,.
\label{eq:cond_indep_criterion_prelim}
\end{equation}
Due to the definition of conditional probability this would require:
\begin{equation}
P(\mathbf{x}_1, \mathbf{y}_3, \mathbf{x}_2, \mathbf{y}_2) = \frac{P(\mathbf{x}_1, \mathbf{x}_2, \mathbf{y}_2) \cdot P(\mathbf{y}_3, \mathbf{x}_2, \mathbf{y}_2)}{P(\mathbf{x}_2, \mathbf{y}_2)}\,.
\label{eq:cond_indep_criterion}
\end{equation}
 
However, marginalizing Eq.~(\ref{eq:non_Markov}) over $\mathbf{x}_3$ and $\mathbf{y}_1$ yields:
\begin{multline}
P(\mathbf{x}_1, \mathbf{y}_3, \mathbf{x}_2, \mathbf{y}_2) \propto \int\!\!\int P(\mathbf{x}_2, \mathbf{y}_1|\mathbf{y}_2, \mathbf{x}_1)\\
\times P(\mathbf{x}_3, \mathbf{y}_2|\mathbf{y}_3, \mathbf{x}_2) \cdot P_{\mathrm{IN}}(\mathbf{x}_1, \mathbf{y}_3)\, d\mathbf{x}_3\, d\mathbf{y}_1\,.\label{eq:joint_from_non_Markov}
\end{multline}
To make the conflict with Eq.~(\ref{eq:cond_indep_criterion}) explicit, define
\begin{equation}
F(x_2,y_2;x_1):=\int P(x_2,y_1\mid y_2,x_1)\,dy_1\,,
\end{equation}
and
\begin{equation}
G(x_2,y_2;y_3):=\int P(x_3,y_2\mid y_3,x_2)\,dx_3\,.
\end{equation}
Then Eq.~(\ref{eq:joint_from_non_Markov}) can be rewritten as
\begin{equation}\label{eq:four_var_joint}
\begin{aligned}
P(x_1,y_3,x_2,y_2)= \frac{1}{Z(x_1,y_3)}\,
P_{\mathrm{IN}}(x_1,y_3)\,\\ \times\, F(x_2,y_2;x_1)\,G(x_2,y_2;y_3)\,,
\end{aligned}
\end{equation}
where
\[Z(x_1,y_3)=\int\!\!\int F(x_2,y_2;x_1)\,G(x_2,y_2;y_3)\,dx_2\,dy_2\]
is the $(x_1,y_3)$-dependent normalization of the conditional bridge propagator. Since $F$ depends on~$x_1$ but not~$y_3$, and $G$ depends on~$y_3$ but not~$x_1$, the conditional distribution
\begin{equation}
\begin{aligned}
P(x_1,y_3\mid x_2,y_2)\propto
\frac{P_{\mathrm{IN}}(x_1,y_3)}{Z(x_1,y_3)}\,\\
 \times\, F(x_2,y_2;x_1)\,G(x_2,y_2;y_3)
\end{aligned}
\end{equation}
factorizes in $x_1$ and $y_3$ \emph{if and only if} $P_{\mathrm{IN}}(x_1,y_3)/Z(x_1,y_3)$ itself factorizes. This makes the origin of the non-Markovianity transparent: the screening-off condition Eq.~(\ref{eq:cond_indep_criterion}) fails whenever the boundary distribution $P_{\mathrm{IN}}(x_1,y_3)$ and the normalization $Z(x_1,y_3)$ do not conspire to produce a product form.
 
That $P_{\mathrm{IN}}(x_1,y_3)$ generically does not factorize seems plausible when one considers the near-classical limit. When the diffusion is weak, the stochastic dynamics approaches deterministic Hamiltonian flow. In this situation, the variables $x$ and $y$ may approximate the canonically conjugate variables $q$ and $p$. For any Hamiltonian with a non-trivial potential $V(q)$, the ``force'' term $-V'(q)$ couples the time evolution of position and momentum, so that the late-time momentum is generically correlated with the early-time position. Since $P_{\mathrm{IN}}(x_1,y_3)$ is the joint distribution over these mixed-time boundary values, it inherits this dynamical coupling and does not factorize. The stochastic case with small but non-zero diffusion is a continuous perturbation of this deterministic limit, so the non-factorization plausibly persists. The process is therefore generically non-Markovian.

\subsection{Bernstein property of the conditioned 
process}\label{subsec:bernstein}
 
While the unconditional dynamics is non-Markovian, 
the process conditioned on boundary data satisfies 
a well-studied structural property that is 
intermediate between full Markovianity and the 
unconstrained non-Markovian behaviour of the 
unconditional process.
 
A stochastic process 
$\{\boldsymbol{\phi}_t\}_{t \in [s, u]}$ 
conditioned on the configurations 
$\boldsymbol{\phi}_s$ and $\boldsymbol{\phi}_u$ 
at the endpoints is said to satisfy the 
\emph{Bernstein property} 
\citep{bernstein1932,jamison1974} if, for any 
$s \leq t_1 < t_2 < t_3 \leq u$, the full 
phase-space configuration 
$\boldsymbol{\phi}_{t_2}$ at the intermediate 
time screens off $\boldsymbol{\phi}_{t_1}$ from 
$\boldsymbol{\phi}_{t_3}$:
\begin{multline}
P(\boldsymbol{\phi}_{t_1}, 
\boldsymbol{\phi}_{t_3} \mid 
\boldsymbol{\phi}_{t_2}, 
\boldsymbol{\phi}_{s}, 
\boldsymbol{\phi}_{u}) \\
= P(\boldsymbol{\phi}_{t_1} \mid 
\boldsymbol{\phi}_{t_2}, 
\boldsymbol{\phi}_{s}) \cdot 
P(\boldsymbol{\phi}_{t_3} \mid 
\boldsymbol{\phi}_{t_2}, 
\boldsymbol{\phi}_{u})\,.
\label{eq:bernstein}
\end{multline}
Equivalently, conditioned on 
$\boldsymbol{\phi}_s$ and 
$\boldsymbol{\phi}_u$, the process on $(s, u)$ 
is Markov.
 
We show in Appendix~\ref{app:bernstein} that the 
process defined by Drummond's path measure 
satisfies the Bernstein property. The key step is 
that the additivity of the action decouples the 
path integral over $[s, t_2]$ from that over 
$[t_2, u]$ once the full configuration 
$\boldsymbol{\phi}_{t_2}$ is fixed, implying the 
conditional independence of the two 
half-trajectories. The extension from the 
mixed-time boundary data 
$(\mathbf{x}_s, \mathbf{y}_u)$ to the full 
endpoint configurations 
$(\boldsymbol{\phi}_{s}, 
\boldsymbol{\phi}_{u})$ then follows from 
standard properties of conditional independence.
 
A further interesting feature is that the distribution of 
$\boldsymbol{\phi}_{t_2}$ at any interior time, 
conditional on the full configurations 
$\boldsymbol{\phi}_{t_1}$ and 
$\boldsymbol{\phi}_{t_3}$ at two bracketing 
times $t_1 < t_2 < t_3$, is independent of the 
boundary data and of the process outside 
$[t_1, t_3]$:
\begin{equation}
P(\boldsymbol{\phi}_{t_2} \mid 
\boldsymbol{\phi}_{t_1}, 
\boldsymbol{\phi}_{t_3}, 
\boldsymbol{\phi}_{s}, 
\boldsymbol{\phi}_{u}) 
= P(\boldsymbol{\phi}_{t_2} \mid 
\boldsymbol{\phi}_{t_1}, 
\boldsymbol{\phi}_{t_3})\,.
\label{eq:interior_shielding}
\end{equation}
Following \citet{bernstein1932}, this is sometimes called the ``reciprocal property'' (or also ``Bernstein property'').

To see why Eq.~(\ref{eq:interior_shielding}) holds, note that the joint probability of 
the configurations at $t_1$, $t_2$, $t_3$ 
factorizes, by additivity of the action, as
\begin{multline}
P(\boldsymbol{\phi}_{t_1}, 
\boldsymbol{\phi}_{t_2}, 
\boldsymbol{\phi}_{t_3} \mid 
\boldsymbol{\phi}_{s}, 
\boldsymbol{\phi}_{u}) \propto 
\mathcal{P}(\boldsymbol{\phi}_{s}, 
\boldsymbol{\phi}_{t_1}) \\
\times\;\mathcal{P}(\boldsymbol{\phi}_{t_1}, 
\boldsymbol{\phi}_{t_2}) \;
\mathcal{P}(\boldsymbol{\phi}_{t_2}, 
\boldsymbol{\phi}_{t_3}) \;
\mathcal{P}(\boldsymbol{\phi}_{t_3}, 
\boldsymbol{\phi}_{u})\,,
\label{eq:five_time_factorization}
\end{multline}
where each factor 
$\mathcal{P}(\boldsymbol{\phi}_{t_a}, 
\boldsymbol{\phi}_{t_b})$ is the result of 
integrating $e^{-S[t_a, t_b]}$ over all paths on 
$[t_a, t_b]$ with the endpoint configurations 
held fixed. For a single time step, such a factor reduces 
to the transition kernel 
$P(\mathbf{x}_{t_b}, \mathbf{y}_{t_a} \mid 
\mathbf{y}_{t_b}, \mathbf{x}_{t_a})$ already 
encountered in Eq.~(\ref{eq:bridge_symm}); for 
multiple steps it is a product of such kernels 
integrated over intermediate configurations. 
Since 
$\mathcal{P}(\boldsymbol{\phi}_{s}, 
\boldsymbol{\phi}_{t_1})$ and 
$\mathcal{P}(\boldsymbol{\phi}_{t_3}, 
\boldsymbol{\phi}_{u})$ do not involve 
$\boldsymbol{\phi}_{t_2}$, they cancel between 
numerator and denominator when one computes, using the definition of conditional probability, 
\begin{multline}
P(\boldsymbol{\phi}_{t_2} \mid 
\boldsymbol{\phi}_{t_1}, 
\boldsymbol{\phi}_{t_3}, 
\boldsymbol{\phi}_{s}, 
\boldsymbol{\phi}_{u})\\
=\frac{P(\boldsymbol{\phi}_{t_1}, 
\boldsymbol{\phi}_{t_2}, 
\boldsymbol{\phi}_{t_3} \mid 
\boldsymbol{\phi}_{s}, 
\boldsymbol{\phi}_{u})}{P(\boldsymbol{\phi}_{t_1}, 
\boldsymbol{\phi}_{t_3} \mid 
\boldsymbol{\phi}_{s}, 
\boldsymbol{\phi}_{u})}\nonumber\,,
\end{multline}
leaving a ratio that 
depends only on $\boldsymbol{\phi}_{t_1}$ and 
$\boldsymbol{\phi}_{t_3}$. This establishes 
Eq.~(\ref{eq:interior_shielding}).

Eq.~(\ref{eq:interior_shielding}) is a physically satisfying property: without it, the probability of finding a given 
configuration at an intermediate time would 
depend on events arbitrarily far in the past or 
future, even when the trajectory is fully pinned 
at nearer times.

Several further remarks are in order.

\emph{(i)} The screening-off requires conditioning 
on the \emph{full} configuration 
$\boldsymbol{\phi}_{t_2} = 
(\mathbf{x}_{t_2}, \mathbf{y}_{t_2})$; 
conditioning on $\mathbf{x}_{t_2}$ or 
$\mathbf{y}_{t_2}$ alone does not suffice, as 
each carries only partial information about the 
state of the process.

\emph{(ii)} The Bernstein property Eq.~(\ref{eq:bernstein}) does \emph{not} restore ordinary (one-sided) Markovianity: the 
factor $P(\boldsymbol{\phi}_{t_3} \mid 
\boldsymbol{\phi}_{t_2}, 
\boldsymbol{\phi}_{u})$ depends on the future 
endpoint $\boldsymbol{\phi}_{u}$, unlike a 
Markov transition kernel. The dynamics therefore 
remains non-Markovian in the sense relevant for 
the no-go theorems discussed in 
Section~\ref{sec:no_go}.

\emph{(iii)} Processes satisfying the Bernstein 
property are closely related to Schr\"odinger 
bridges \citep{schrodinger1931,leonard2014}. In 
that literature, ``reversible'' Brownian motion 
refers to a standard (time-asymmetric) diffusion 
conditioned on both-endpoint data; the resulting 
bridge satisfies the Bernstein property, with transition probabilities that are symmetric between the two time directions. In 
the present framework, by contrast, the time 
symmetry is built into the dynamics itself through 
the traceless diffusion matrix, prior to any 
conditioning. The two constructions yield 
structurally similar objects via different routes; 
whether a formal connection can be established is 
an interesting question for future work 
\citep{zambrini1986}.

\subsection{Physical interpretation}

In the light of the argument based on an observation by Watanabe presented in Section~2 of the companion paper \citep{friederichtyagi_PartI}, it is unsurprising that combining stochasticity with time-reversal invariance does not lead to dynamics governed by a law-like Markov kernel. In that sense, the breakdown of Markovianity in Drummond's dynamics as demonstrated here is only to be expected. As Adlam \citep{adlam2018} has argued independently, some form of temporal non-locality, which can have the form of non-Markovian dynamics, may actually be seen as natural alongside the spatial non-locality implied by Bell inequality violations. More recently, Barandes \citep{barandes2023} has argued that violations of the Markov condition are essential for recovering quantum theory from stochastic dynamics.

The time-symmetric dynamics fits naturally with the block universe view of modern physics advocated perhaps most famously by Price \citep{price1996}: trajectories exist in spacetime as four-dimensional worldlines, and the stochastic laws describe constraints on these worldlines rather than recipes for constructing the future from the past.

\section{Relation to no-go theorems}\label{sec:no_go}

The non-Markovian structure of the trajectory 
dynamics has important implications for how this approach relates to no-go theorems in quantum foundations. As we now show, the most well-known such theorems do not constrain the approach. The real 
obstacles are the representability gap 
(Section~\ref{subsec:representability_gap}) and 
the restriction to Hamiltonians with at most 
quartic density-density interactions.

\subsection{Failure of 
$\lambda$-mediation}\label{subsec:lambda_med}

Many influential no-go theorems are formulated 
within the ontological models framework 
\citep{harrigan2010}, a central assumption of 
which is \emph{$\lambda$-mediation}: measurement 
outcome probabilities conditional on the ontic 
state $\lambda$ are independent of the 
preparation procedure $R$. In the present 
framework, the natural candidate for $\lambda$ is 
the full phase-space configuration 
$\boldsymbol{\phi}_t = (\mathbf{x}_t, 
\mathbf{y}_t)$ at some intermediate time $t$ 
between preparation and measurement (or over a 
time interval). We now show that 
$\lambda$-mediation fails for this identification.

The key observation is that the time-oriented 
conditional probability 
$P_R(\boldsymbol{\phi}_{t_2} \mid 
\boldsymbol{\phi}_{t_1})$, which governs the 
evolution from an earlier to a later phase-space 
configuration, depends on the preparation $R$. 
This can be seen directly from Bayes' theorem, taking into account that the mixed-time conditional probability 
$P(\mathbf{x}_2, \mathbf{y}_1 \mid \mathbf{y}_2, 
\mathbf{x}_1)$ -- the propagator defined by the 
traceless diffusion Fokker-Planck equation -- is 
law-like: it is fixed entirely by the Hamiltonian 
and does not depend on the preparation (which is why it does not carry an index $R$. The time-oriented conditional $P_R(\boldsymbol{\phi}_2 \mid 
\boldsymbol{\phi}_1)$ relates to this propagator via 
Bayes' theorem as
\begin{equation}
P_R(\boldsymbol{\phi}_2 \mid 
\boldsymbol{\phi}_1) = 
P(\mathbf{x}_2, \mathbf{y}_1 \mid 
\mathbf{y}_2, \mathbf{x}_1) \cdot 
\frac{P_R(\mathbf{y}_2, \mathbf{x}_1)}
{P_R(\boldsymbol{\phi}_1)}\,,
\label{eq:bayes_violation}
\end{equation}
where $\boldsymbol{\phi}_k = (\mathbf{x}_k, 
\mathbf{y}_k)$, and $P_R(\mathbf{y}_2, 
\mathbf{x}_1)$ and $P_R(\boldsymbol{\phi}_1)$ are marginal distributions that 
depend on the $Q$-function and hence on the 
preparation $R$ in a non-trivial way. Since the time-oriented 
conditional inherits this dependence, measurement 
outcome probabilities conditional on 
$\boldsymbol{\phi}_t$ are not independent of 
$R$: $\lambda$-mediation fails.

This failure is not an ad hoc feature of the 
present framework. As shown in Section~2 of our 
companion paper \citep{friederichtyagi_PartI}, 
it follows from an observation by Watanabe \citep{watanabe1965} 
that in any stochastic theory, if the 
time-oriented conditional 
$P(\boldsymbol{\phi}_2 \mid 
\boldsymbol{\phi}_1)$ were law-like, then by 
Bayes' theorem the reverse conditional 
$P(\boldsymbol{\phi}_1 \mid 
\boldsymbol{\phi}_2)$ would depend on the 
preparation -- breaking time-reversal invariance. 
The only way for both time-oriented conditionals 
to be simultaneously law-like would be for the 
unconditional probabilities to be law-like as 
well, which, in the present approach, would mean embracing an ``ontic'' interpretation of the $Q$-function and thereby abandoning the conceptual analogy with classical statistical mechanics. In a genuinely time-symmetric stochastic theory, therefore, 
neither time-oriented conditional can be law-like, 
and one should expect $\lambda$-mediation to fail.

In what follows we sketch out how the failure of $\lambda$-mediation makes different kinds of no-go theorems inapplicable to the approach based on Drummond's dynamics.

\subsection{Consequences for specific 
theorems}\label{subsec:consequences}

\subsubsection{$\psi$-ontology theorems}
The Pusey-Barrett-Rudolph (PBR) theorem 
\citep{pusey2012} establishes that, within the 
ontological models framework, distinct quantum 
states must correspond to non-overlapping 
probability distributions over ontic states 
$\lambda$. Related results by Hardy 
\citep{hardy2013} and Colbeck-Renner 
\citep{colbeck2017} reach similar conclusions. 
These theorems would rule out an epistemic 
interpretation of the quantum state: if different 
$Q$-functions had to correspond to non-overlapping 
distributions over phase-space configurations 
$\boldsymbol{\phi}$, the $Q$-function could not 
be viewed as representing incomplete knowledge of 
an underlying reality.

All of these results, however, rely on 
$\lambda$-mediation \citep{leifer2014}: they 
assume that the probability of a measurement 
outcome, conditional on the ontic state 
$\lambda$, does not depend on the preparation. As 
shown in 
Section~\ref{subsec:lambda_med}, this assumption 
fails in the present framework because the 
time-oriented conditional 
$P_R(\boldsymbol{\phi}_2 \mid 
\boldsymbol{\phi}_1)$ is preparation-dependent. 
The theorems therefore do not apply, and Husimi 
functions for different quantum states are free to 
overlap. This leaves open a genuinely epistemic 
interpretation of the quantum state: the 
$Q$-function can be understood as encoding 
incomplete knowledge of the actual phase-space 
configuration, much as the probability density in 
classical statistical mechanics encodes incomplete 
knowledge of the actual microstate.

\subsubsection{Spekkens contextuality and negativity results}

At first glance, several results appear to pose serious challenges to the approach developed here. Spekkens \citep{spekkens2005} demonstrated that any model reproducing quantum predictions must exhibit preparation contextuality and measurement contextuality. This means that in such a model, the preparation procedure $R$ cannot be simply identified with the prepared quantum state $\hat{\rho}$, and the measurement procedure $M$ cannot be simply identified with the measured POVM $\{\hat{E}_m\}$ -- the probabilities must depend on additional contextual information beyond what is captured by $\hat{\rho}$ and $\{\hat{E}_m\}$ alone. These contextual dependencies appear prima facie unattractive, seemingly undermining the naturalness of hidden variable theories.

Even more strikingly, Spekkens' results on negativity \citep{spekkens2008} and related work by Ferrie and Emerson \citep{ferrie2008} appear to show the outright \emph{impossibility} of reproducing quantum predictions using non-negative quasi-probability distributions such as the Husimi function. Specifically, these results establish that in any model reproducing the Born rule
\begin{equation}
\mathrm{Tr}(\hat{E}_m\hat{\rho}) = \int P_M(a_m|\lambda) P_R(\lambda) d\lambda\,,\label{eq:ontological_models}
\end{equation}
at least one of the probability distributions $P_R(\lambda)$ (over ontic states given preparation procedure $R$) or $P_M(a_m|\lambda)$ (over measurement outcomes $a_m$ given ontic state $\lambda$ and measurement procedure $M$) must sometimes take negative values.

However, all these results crucially assume $\lambda$-mediation: that the conditional probabilities $P_M(a_m|\lambda)$ of measurement outcomes given the ontic state are independent of the preparation procedure $R$. As we have seen, this assumption fails in the present approach -- measurement outcome probabilities conditional on phase space location \emph{do} depend on the preparation procedure. Consequently, none of these no-go theorems directly constrain the present framework.

For the contextuality results, this means that, indeed, the approach proposed here does not recover quantum predictions as weighted integrals over some ontic variable $\lambda$ along the lines of Eq.~(\ref{eq:ontological_models}) and in that sense it is ``contextual.'' However, this ``contextuality'' did not arise from contrived or ad hoc assumptions designed to evade the no-go theorem. Rather, it follows naturally and necessarily from the time-symmetric, non-Markovian structure of the dynamics, as established in Section~2 of the companion paper and in Section~\ref{sec:non-Markovian}. The dependence of outcome probabilities on more than just the ontic state at measurement time is simply a manifestation of the backward-temporal dependencies encoded in the mixed-time boundary conditions that define the theory's trajectories.

For the negativity results of Spekkens and Ferrie-Emerson, the situation is even simpler: These theorems establish impossibility of representing quantum states with a positive probability density within the ontological models framework, which assumes $\lambda$-mediation. Since the present approach violates it, the approach lies outside the scope of these impossibility proofs. The Husimi function is non-negative and normalized to unity, providing a genuine probability distribution over phase space. No quasi-probability with negative values is required, despite what the negativity theorems might suggest at first glance. The resolution is that the theorems' assumptions -- specifically, the separation of preparation and measurement contexts in a way that respects $\lambda$-mediation -- simply do not hold in time-symmetric stochastic theories.

\subsubsection{Bell's theorem} 
Bell's theorem \citep{bell1964} establishes that the correlations implied by quantum theory cannot be reproduced in theories that respect \emph{statistical independence} (variables in the common past of two measurements far apart from each other do not correlate with the later measurement settings) and \emph{local causality} (conditional on the variables in the common past, distant outcomes do not correlate with each other nor with distant measurement settings).
 
In the present approach, \emph{statistical independence} is preserved. Phase space location is the ontic variable, and the probability density over phase space location at any given time -- the Husimi function, equivalent to the quantum state -- is assigned based on data about past (or present), not future, observations, hence there is no correlation between ontic variables and later measurement settings. Since the interpretation of the Husimi function as a proper probability density can account for Bell inequality violations, as shown by Reid and Drummond \citep{drummondreid2026}, \emph{local causality} must be violated.
 
Reid and Drummond \citep{drummondreid2026} consider a model of Bell inequality violations in the forward-backward stochastic framework for continuous-variable systems. In that model, measurement of a one-mode system is modelled as parametric amplification: phase space coordinates $q$ and $p$ correspond to mode quadratures; one quadrature -- the ``measured'' one -- is exponentially amplified while the conjugate quadrature is attenuated. Measurement outcomes are determined by the amplified quadrature, which requires marginalising over the attenuated one. The $Q$-function of an entangled state contains off-diagonal interference terms that oscillate in the deamplified variables; this marginalisation washes them out, so that with fixed measurement settings the observed statistics are those of a classical mixture. A local change of measurement setting rotates phase-space coordinates, partially mixing the amplified and deamplified quadratures. Reid and Drummond show that a rotation at a single site is not sufficient to make the interference terms survive marginalisation, but that rotations at \emph{both} sites can, producing joint probabilities that violate Bell inequalities.
 
How does this violation relate to the time-symmetric causal structure? In the forward-backward dynamics, the amplified quadratures propagate backward from a future boundary condition, while the deamplified quadratures propagate forward from a past boundary. The $Q$~function at the preparation time links these two streams. For an entangled state, this linking is non-factorizable between the two subsystems. If one considers a broader ensemble with uncorrelated future boundary values -- not corresponding to any particular quantum state -- then selecting the sub-ensemble whose early-time configuration matches a specific entangled $Q$~function induces correlations among the future boundary values. State preparation, viewed time-symmetrically, amounts to precisely such a sub-ensemble selection. This perspective can be related to a recent proposal by Price and Wharton \citep{price2021,price2024} according to which conditioning on a past ``collider'' (see \citep{elwertwinship2014} for a general introduction to this phenomenon) can generate Bell-violating correlations. Concrete models in which, in the presence of past-directed law-like conditional probabilistic dependencies, conditioning on a past ``collider'' recovers the quantum predictions for Bell and GHZ states are developed in \citep{friederich2024eprl}. In the amplification model that exhibits Bell inequality violations, however, the conditioning is not on a single phase-space value but on the full early-time $Q$~function -- an ensemble-level constraint. Whether this distributional conditioning admits a precise causal-graphical interpretation remains an open question.

\section{Summary and outlook}\label{sec:summary}

In this paper we considered whether a trajectory interpretation is available for those quantum field theories whose time evolution equation for the Husimi function reduces to the Fokker-Planck equation with traceless diffusion motivated in the companion paper. Such a trajectory interpretation is attractive because it allows one to interpret those quantum field theories as conceptually analogous to classical statistical mechanics, though with an underlying stochastic dynamics. 

The findings of Drummond in \citep{drummond2021} show that one can indeed provide a measure over trajectories that leads to precisely this evolution equation for the phase space probability density. However, as acknowledged by Drummond, those findings do not establish whether actual Husimi functions can always be represented in this way -- we have called this lacuna the ``representability gap'' (Section~\ref{subsec:representability_gap}). As long as this gap remains open, it is unclear whether a trajectory interpretation along the lines constructed by Drummond is indeed available for the quantum field theories in question.

We then proceeded to investigate the dynamical features of Drummond's phase space trajectories, the potential limits to their applicability notwithstanding. It turns out that Drummond's measure over paths gives rise to non-Markovian dynamics, while the so-called Bernstein property is respected.

The failure of Markovianity has an interesting consequence, namely, it gives rise to the failure of $\lambda$-mediation, which is a seemingly plausible key assumption of the widely used ontological models framework. In the absence of $\lambda$-mediation, the key no-go theorems on ``hidden variable'' interpretations can no longer be derived, which makes it possible to reconcile the empirical success of the quantum field theories in question with attributing to the system a sharp yet unknown phase space location. It should be noted that the plausibility of $\lambda$-mediation notwithstanding, in Drummond's dynamics its failure is not an ad hoc manoeuvre, but a consequence of the measure over phase space paths considered on independent grounds. 

The moral of these observations, as we see it, is that the real challenges to providing a compelling theoretical approach that permits us to view quantum field theories as the statistical mechanics of underlying time-symmetric stochastic processes do not come from the no-go theorems. Instead, the real challenge lies in closing the representability gap 
(Section~\ref{subsec:representability_gap}) and ultimately 
in extending a future development of the present framework to cover all quantum field theories for which we have empirical evidence.

\appendix
 
\section{Proof of the Bernstein 
property}\label{app:bernstein}
 
We prove that the process defined by Drummond's 
path measure satisfies the Bernstein property 
stated in Section~\ref{subsec:bernstein}. The 
argument proceeds in two steps: first, we 
establish \emph{path-level independence} of the 
two half-intervals from the additivity of the 
action; second, we extend the result from the 
mixed-time boundary data to full endpoint 
conditioning using standard semi-graphoid axioms.
 
\subsection{Path-level independence and the 
mixed-time Bernstein 
property}\label{app:path_level}
 
Consider three intermediate times 
$s < t_1 < t_2 < t_3 < u$. By the 
additivity of the action,
\begin{equation}
S[\boldsymbol{\phi}]_{s}^{u} 
= S[\boldsymbol{\phi}]_{s}^{t_2} 
+ S[\boldsymbol{\phi}]_{t_2}^{u}\,,
\end{equation}
each piece depends only on the trajectory within 
its own time interval. Fixing the full 
configuration $\boldsymbol{\phi}_{t_2}$ at the 
intermediate time therefore decouples the path 
integral over $[s, t_2]$ from that over 
$[t_2, u]$, implying the independence of the 
\emph{entire trajectory} on $[s, t_2]$ from 
the \emph{entire trajectory} on $[t_2, u]$, 
conditional on 
$(\boldsymbol{\phi}_{t_2}, \mathbf{x}_s, 
\mathbf{y}_u)$.
 
In particular, the joint density of 
configurations at $t_1$, $t_2$, and $t_3$ 
factorises:
\begin{multline}
P(\boldsymbol{\phi}_{t_1}, 
\boldsymbol{\phi}_{t_2}, 
\boldsymbol{\phi}_{t_3} \mid \mathbf{x}_s, 
\mathbf{y}_u) \\
\propto 
F(\boldsymbol{\phi}_{t_1}, 
\boldsymbol{\phi}_{t_2}; \mathbf{x}_s) \cdot 
G(\boldsymbol{\phi}_{t_2}, 
\boldsymbol{\phi}_{t_3}; \mathbf{y}_u)\,,
\label{eq:FG_factorisation}
\end{multline}
where $F$ is the path integral over 
$[s, t_2]$ with boundary datum $\mathbf{x}_s$ 
and intermediate configurations 
$\boldsymbol{\phi}_{t_1}$, 
$\boldsymbol{\phi}_{t_2}$ pinned, and $G$ is 
the corresponding integral over $[t_2, u]$ 
with datum $\mathbf{y}_u$ and configurations 
$\boldsymbol{\phi}_{t_2}$, 
$\boldsymbol{\phi}_{t_3}$ pinned. Crucially, 
$F$ depends on $\mathbf{x}_s$ but not on 
$\mathbf{y}_u$, and $G$ depends on 
$\mathbf{y}_u$ but not on $\mathbf{x}_s$.
 
Defining marginalisation factors 
$\bar{F}(\boldsymbol{\phi}_{t_2}; 
\mathbf{x}_s) = \int F\, 
d\boldsymbol{\phi}_{t_1}$ and 
$\bar{G}(\boldsymbol{\phi}_{t_2}; 
\mathbf{y}_u) = \int G\, 
d\boldsymbol{\phi}_{t_3}$, the conditional 
density is:
\begin{multline}
P(\boldsymbol{\phi}_{t_1}, 
\boldsymbol{\phi}_{t_3} \mid 
\boldsymbol{\phi}_{t_2}, \mathbf{x}_s, 
\mathbf{y}_u) \\
= 
\frac{F(\boldsymbol{\phi}_{t_1}, 
\boldsymbol{\phi}_{t_2}; \mathbf{x}_s)}
{\bar{F}(\boldsymbol{\phi}_{t_2}; 
\mathbf{x}_s)} \cdot 
\frac{G(\boldsymbol{\phi}_{t_2}, 
\boldsymbol{\phi}_{t_3}; \mathbf{y}_u)}
{\bar{G}(\boldsymbol{\phi}_{t_2}; 
\mathbf{y}_u)}\,,
\label{eq:bernstein_mixed}
\end{multline}
where each ratio is separately normalised. Since the right-hand side factorizes into a function of $(\boldsymbol{\phi}_{t_1}, \boldsymbol{\phi}_{t_2}, \mathbf{x}_s)$ times a function of $(\boldsymbol{\phi}_{t_2}, \boldsymbol{\phi}_{t_3}, \mathbf{y}_u)$, this is exactly the statement that 
\begin{equation}
    \boldsymbol{\phi}_{t_1} 
\;\perp\!\!\!\perp\; 
\boldsymbol{\phi}_{t_3} 
\;\mid\; 
(\boldsymbol{\phi}_{t_2}, 
\mathbf{x}_s, 
\mathbf{y}_u)\,.
\end{equation}
This establishes the Bernstein property conditional on 
the mixed-time boundary data 
$(\mathbf{x}_s, \mathbf{y}_u)$.

\subsection{Extension to full endpoint 
conditioning}\label{app:full_endpoint}
 
We now show that the conditional independence of 
$\boldsymbol{\phi}_{t_1}$ and 
$\boldsymbol{\phi}_{t_3}$ is preserved when one 
conditions additionally on the ``output'' 
boundary values $\mathbf{y}_s = \mathbf{y}(s)$ 
and $\mathbf{x}_u = \mathbf{x}(u)$, thereby 
establishing Eq.~(\ref{eq:bernstein}). The 
argument uses a property of conditional 
independence that holds in any probability distribution, which is known as \textit{weak union}
\citep{dawid1979,pearl1988}. Namely, for disjoint sets 
of random variables $X$, $Y$, $W$, and $Z$:
 
\begin{description}
\item[Weak union.] If $X$ is conditionally 
independent of $Y$ and $W$ jointly, given $Z$, 
then $X$ is conditionally independent of $Y$, 
given $Z$ and $W$:
\begin{equation}
X \perp\!\!\!\perp (Y, W) \mid Z 
\;\;\Longrightarrow\;\; 
X \perp\!\!\!\perp Y \mid (Z, W)\,.
\label{eq:weak_union}
\end{equation}
\end{description}
 
\noindent In addition, we will use that conditional independence is symmetric in that $X \perp\!\!\!\perp Y \mid Z$ if and only if  $Y \perp\!\!\!\perp X \mid Z$.
 
The path-level independence established in 
Section~\ref{app:path_level} implies that, 
conditional on 
$C \coloneqq (\boldsymbol{\phi}_{t_2}, 
\mathbf{x}_s, \mathbf{y}_u)$, all functionals 
of the trajectory on $[s, t_2]$ are jointly 
independent of all functionals of the trajectory 
on $[t_2, u]$. Since 
$\boldsymbol{\phi}_{t_1}$ and $\mathbf{y}_s$ 
are functionals of the first half-trajectory, 
and $\boldsymbol{\phi}_{t_3}$ and $\mathbf{x}_u$ 
are functionals of the second, the conditional independence of the two half-trajectories implies the conditional independence of these derived random variables:
\begin{equation}
(\boldsymbol{\phi}_{t_1}, \mathbf{y}_s) 
\;\perp\!\!\!\perp\; 
(\boldsymbol{\phi}_{t_3}, \mathbf{x}_u) 
\;\mid\; C\,.
\label{eq:path_CI}
\end{equation}
 
The derivation of Eq.~(\ref{eq:bernstein}) 
proceeds in three steps.
 
\emph{Step 1 (weak union).} Apply weak union to 
Eq.~(\ref{eq:path_CI}), moving $\mathbf{x}_u$ 
from the right-hand side into the conditioning 
set:
\begin{equation}
(\boldsymbol{\phi}_{t_1}, \mathbf{y}_s) 
\;\perp\!\!\!\perp\; 
\boldsymbol{\phi}_{t_3} 
\;\mid\; (C, \mathbf{x}_u)\,.
\label{eq:step1}
\end{equation}
 
\emph{Step 2 (symmetry and weak union).} By 
symmetry, Eq.~(\ref{eq:step1}) is equivalent to
\[
\boldsymbol{\phi}_{t_3} 
\;\perp\!\!\!\perp\; 
(\boldsymbol{\phi}_{t_1}, \mathbf{y}_s) 
\;\mid\; (C, \mathbf{x}_u)\,.
\]
Applying weak union, moving $\mathbf{y}_s$ into 
the conditioning set:
\begin{equation}
\boldsymbol{\phi}_{t_3} 
\;\perp\!\!\!\perp\; 
\boldsymbol{\phi}_{t_1} 
\;\mid\; (C, \mathbf{x}_u, \mathbf{y}_s)\,.
\label{eq:step2}
\end{equation}
 
\emph{Step 3 (symmetry).} Applying symmetry to 
Eq.~(\ref{eq:step2}) and noting that 
$(C, \mathbf{x}_u, \mathbf{y}_s) = 
(\boldsymbol{\phi}_{t_2}, 
\boldsymbol{\phi}_{s}, 
\boldsymbol{\phi}_{u})$:
\begin{equation}
\boldsymbol{\phi}_{t_1} 
\;\perp\!\!\!\perp\; 
\boldsymbol{\phi}_{t_3} 
\;\mid\; 
(\boldsymbol{\phi}_{t_2}, 
\boldsymbol{\phi}_{s}, 
\boldsymbol{\phi}_{u})\,,
\label{eq:bernstein_full}
\end{equation}
which is the Bernstein property 
Eq.~(\ref{eq:bernstein}), for the endpoint-conditioned process. \qed

\vspace{1cm}
 
\begin{acknowledgments}
We would like to thank Benjamin Feintzeig for helpful feedback of an earlier version. We used Anthropic's Claude Opus 4.5 and 4.6 for feedback on our ideas and for assistance with working out the arguments presented here. We take responsibility for everything in the paper, all potentially remaining errors are ours.

This research was funded by the Netherlands Organization for Scientific Research (NWO), project VI.Vidi.211.088.

\end{acknowledgments}

\bibliographystyle{apsrev4-2}
\bibliography{bibliography}

@article{adlam2018,
  author = {Adlam, Emily},
  title = {Spooky Action at a Temporal Distance},
  journal = {Entropy},
  volume = {20},
  pages = {41},
  year = {2018},
  doi = {10.3390/e20010041}
}

@misc{appleby1998optimaljointmeasurementsposition,
      title={Optimal Joint Measurements of Position and Momentum}, 
      author={D. M. Appleby},
      year={1998},
      eprint={quant-ph/9803053},
      archivePrefix={arXiv},
      primaryClass={quant-ph},
      url={https://arxiv.org/abs/quant-ph/9803053}, 
}

@article{barandes2023,
  author = {Barandes, Jacob A.},
  title = {The stochastic-quantum correspondence},
  journal = {Philosophy of Physics},
  year = {2025},
  volume = {3},
  pages = {8},
  doi = {10.31389/pop.186}
}

@article{bell1964,
  author = {Bell, John S.},
  title = {On the Einstein-Podolsky-Rosen paradox},
  journal = {Physics},
  volume = {1},
  pages = {195--200},
  year = {1964}
}

@article{bopp1956,
  author = {Bopp, Fritz},
  title = {La m{\'e}canique quantique est-elle une m{\'e}canique statistique classique particuli{\`e}re?},
  journal = {Annales de l'Institut Henri Poincar{\'e}},
  volume = {15},
  pages = {81--112},
  year = {1956}
}

@article{drummond2021,
  author = {Drummond, Peter D.},
  title = {Time evolution with symmetric stochastic action},
  journal = {Physical Review Research},
  volume = {3},
  pages = {013240},
  year = {2021},
  doi = {10.1103/PhysRevResearch.3.013240}
}

@article{drummondreid2020,
  author = {Drummond, Peter D. and Reid, Margaret D.},
  title = {Retrocausal model of reality for quantum fields},
  journal = {Physical Review Research},
  volume = {2},
  pages = {033266},
  year = {2020},
  doi = {10.1103/PhysRevResearch.2.033266}
}

@article{drummondreid2026,
  author = {Reid, Margaret D. and Drummond, Peter D.},
  title = {Forward-backward stochastic simulations: {Q}-based model for measurement and {B}ell nonlocality consistent with weak local realistic premises},
  journal = {Physical Review A},
  volume = {113},
  pages = {012210},
  year = {2026},
  doi = {10.1103/PhysRevA.113.012210}
}

@article{elwertwinship2014,
  author = {Elwert, Felix and Winship, Christopher},
  title = {Endogenous selection bias: The problem of conditioning on a collider variable},
  journal = {Annual Review of Sociology},
  volume = {40},
  pages = {31--53},
  year = {2014},
  doi = {10.1146/annurev-soc-071913-043455}
}

@article{ferrie2008,
  author = {Ferrie, Christopher and Emerson, Joseph},
  title = {Frame representations of quantum mechanics and the necessity of negativity in quasi-probability representations},
  journal = {Journal of Physics A: Mathematical and Theoretical},
  volume = {41},
  pages = {352001},
  year = {2008},
  doi = {10.1088/1751-8113/41/35/352001}
}

@book{folland1989,
  author = {Folland, Gerald B.},
  title = {Harmonic Analysis in Phase Space},
  publisher = {Princeton University Press},
  address = {Princeton},
  year = {1989}
}

@article{friederich2024,
  author = {Friederich, Simon},
  title = {Introducing the {Q}-based interpretation of quantum mechanics},
  journal = {British Journal for the Philosophy of Science},
  volume = {75},
  pages = {769--795},
  year = {2024},
  doi = {10.1086/714810}
}

@article{friederich2024eprl,
  author = {Friederich, Simon},
  title = {Reproducing {EPR} correlations without superluminal signalling: Backward conditional probabilities and Statistical Independence},
  journal = {Europhysics Letters},
  volume = {150},
  pages = {40001},
  year = {2025},
  doi = {10.1209/0295-5075/ad8c00}
}

@article{depolavieja1996,
title = {A causal quantum theory in phase space},
journal = {Physics Letters A},
volume = {220},
number = {6},
pages = {303-314},
year = {1996},
issn = {0375-9601},
doi = {https://doi.org/10.1016/0375-9601(96)00523-3},
url = {https://www.sciencedirect.com/science/article/pii/0375960196005233},
author = {Gonzalo {García de Polavieja}}
}

@article{graham1977,
  author = {Graham, Robert},
  title = {Path integral formulation of general diffusion processes},
  journal = {Zeitschrift f{\"u}r Physik B},
  volume = {26},
  pages = {281--290},
  year = {1977},
  doi = {10.1007/BF01312935}
}

@article{harrigan2010,
  author = {Harrigan, Nicholas and Spekkens, Robert W.},
  title = {Einstein, incompleteness, and the epistemic view of quantum states},
  journal = {Foundations of Physics},
  volume = {40},
  pages = {125--157},
  year = {2010},
  doi = {10.1007/s10701-009-9347-0}
}

@article{husimi1940,
  author = {Husimi, K{\^o}di},
  title = {Some formal properties of the density matrix},
  journal = {Proceedings of the Physico-Mathematical Society of Japan},
  volume = {22},
  pages = {264--314},
  year = {1940}
}

@article{kochen1967,
  author = {Kochen, Simon and Specker, Ernst P.},
  title = {The problem of hidden variables in quantum mechanics},
  journal = {Journal of Mathematics and Mechanics},
  volume = {17},
  pages = {59--87},
  year = {1967}
}

@article{lee,
title = {Theory and application of the quantum phase-space distribution functions},
journal = {Physics Reports},
volume = {259},
number = {3},
pages = {147-211},
year = {1995},
issn = {0370-1573},
doi = {https://doi.org/10.1016/0370-1573(95)00007-4},
url = {https://www.sciencedirect.com/science/article/pii/0370157395000074},
author = {Hai-Woong Lee}
}

@article{leifer2014,
  author = {Leifer, Matthew S.},
  title = {Is the quantum state real? {A}n extended review of $\psi$-ontology theorems},
  journal = {Quanta},
  volume = {3},
  pages = {67--155},
  year = {2014},
  doi = {10.12743/quanta.v3i1.22}
}

@article{miranker1961,
  author = {Miranker, Willard L.},
  title = {A well posed problem for the backward heat equation},
  journal = {Proceedings of the American Mathematical Society},
  volume = {12},
  pages = {243--247},
  year = {1961},
  doi = {10.1090/S0002-9939-1961-0120462-2}
}

@article{PhysRevLett.117.070801,
  title = {Evading Vacuum Noise: Wigner Projections or Husimi Samples?},
  author = {M\"uller, C. R. and Peuntinger, C. and Dirmeier, T. and Khan, I. and Vogl, U. and Marquardt, Ch. and Leuchs, G. and S\'anchez-Soto, L. L. and Teo, Y. S. and Hradil, Z. and \ifmmode \check{R}\else \v{R}\fi{}eh\'a\ifmmode \check{c}\else \v{c}\fi{}ek, J.},
  journal = {Phys. Rev. Lett.},
  volume = {117},
  issue = {7},
  pages = {070801},
  numpages = {6},
  year = {2016},
  month = {Aug},
  publisher = {American Physical Society},
  doi = {10.1103/PhysRevLett.117.070801},
  url = {https://link.aps.org/doi/10.1103/PhysRevLett.117.070801}
}

@article{onsagermachlup1953,
  author = {Onsager, Lars and Machlup, Stefan},
  title = {Fluctuations and irreversible processes},
  journal = {Physical Review},
  volume = {91},
  number = {6},
  pages = {1505--1512},
  year = {1953},
  doi = {10.1103/PhysRev.91.1505}
}

@book{price1996,
  author = {Price, Huw},
  title = {Time's Arrow and {A}rchimedes' Point: New Directions for the Physics of Time},
  publisher = {Oxford University Press},
  year = {1996}
}

@article{price2021,
  author = {Price, Huw and Wharton, Ken},
  title = {Entanglement swapping and action at a distance},
  journal = {Foundations of Physics},
  volume = {51},
  pages = {105},
  year = {2021},
  doi = {10.1007/s10701-021-00512-5}
}

@article{price2024,
  author = {Price, Huw and Wharton, Ken},
  title = {A mechanism for entanglement?},
  journal = {arXiv preprint arXiv:2406.04571},
  year = {2024},
  eprint = {2406.04571},
  archivePrefix = {arXiv}
}

@article{pusey2012,
  author = {Pusey, Matthew F. and Barrett, Jonathan and Rudolph, Terry},
  title = {On the reality of the quantum state},
  journal = {Nature Physics},
  volume = {8},
  pages = {475--478},
  year = {2012},
  doi = {10.1038/nphys2309}
}

@article{spekkens2005,
  author = {Spekkens, Robert W.},
  title = {Contextuality for preparations, transformations, and unsharp measurements},
  journal = {Physical Review A},
  volume = {71},
  pages = {052108},
  year = {2005},
  doi = {10.1103/PhysRevA.71.052108}
}

@article{spekkens2008,
  author = {Spekkens, Robert W.},
  title = {Negativity and contextuality are equivalent notions of nonclassicality},
  journal = {Physical Review Letters},
  volume = {101},
  pages = {020401},
  year = {2008},
  doi = {10.1103/PhysRevLett.101.020401}
}

@article{tyagifriederich,
  author = {Tyagi, Mritunjay and Friederich, Simon},
  title = {Time evolution of the {H}usimi and {G}lauber-{S}udarshan functions in terms of complementary {H}amiltonian symbols},
  journal = {arXiv preprint arXiv:2510.15628},
  year = {2025},
  note = {Preprint},
  eprint = {},
  archivePrefix = {},
  primaryClass = {quant-ph}
}

@article{watanabe1965,
  author = {Watanabe, Satosi},
  title = {Conditional probability in physics},
  journal = {Progress of Theoretical Physics Supplement},
  volume = {E65},
  pages = {135--160},
  year = {1965},
  note = {Supplement dedicated to Yuzuru Watanabe}
}

@book{Wiseman_Milburn_2009, place={Cambridge}, title={Quantum Measurement and Control}, publisher={Cambridge University Press}, author={Wiseman, Howard M. and Milburn, Gerard J.}, year={2009}}

@article{bernstein1932,
  author  = {S. Bernstein},
  title   = {Sur les liaisons entre les grandeurs al\'eatoires},
  journal = {Verh. Int. Math.-Kongr. Z\"urich},
  volume  = {1},
  pages   = {288--309},
  year    = {1932}
}

@article{zambrini1986,
  author  = {J. C. Zambrini},
  title   = {Variational processes and stochastic versions of mechanics},
  journal = {J. Math. Phys.},
  volume  = {27},
  pages   = {2307--2330},
  year    = {1986}
}

@article{schrodinger1931,
  author  = {E. Schr{\"o}dinger},
  title   = {{\"U}ber die {U}mkehrung der {N}aturgesetze},
  journal = {Sitzungsberichte der Preussischen Akademie der 
             Wissenschaften, Physikalisch-Mathematische Klasse},
  volume  = {9},
  pages   = {144--153},
  year    = {1931}
}

@article{leonard2014,
  author  = {C. L{\'e}onard},
  title   = {A survey of the {S}chr{\"o}dinger problem and some of 
             its connections with optimal transport},
  journal = {Discrete and Continuous Dynamical Systems},
  volume  = {34},
  number  = {4},
  pages   = {1533--1574},
  year    = {2014},
  doi     = {10.3934/dcds.2014.34.1533}
}

@article{dawid1979,
  author  = {A. P. Dawid},
  title   = {Conditional independence in statistical 
             theory},
  journal = {Journal of the Royal Statistical 
             Society, Series B},
  volume  = {41},
  number  = {1},
  pages   = {1--31},
  year    = {1979}
}

@book{pearl1988,
  author    = {J. Pearl},
  title     = {Probabilistic Reasoning in Intelligent 
               Systems: Networks of Plausible 
               Inference},
  publisher = {Morgan Kaufmann},
  address   = {San Mateo, CA},
  year      = {1988}
}

@article{jamison1974,
  author  = {B. Jamison},
  title   = {Reciprocal processes},
  journal = {Zeitschrift f\"ur 
             Wahrscheinlichkeitstheorie und 
             verwandte Gebiete},
  volume  = {30},
  pages   = {65--86},
  year    = {1974},
  doi     = {10.1007/BF00532864}
}

@article{hardy2013,
  author  = {L. Hardy},
  title   = {Are quantum states real?},
  journal = {International Journal of Modern 
             Physics B},
  volume  = {27},
  pages   = {1345012},
  year    = {2013},
  doi     = {10.1142/S0217979213450124}
}

@article{colbeck2017,
  author  = {R. Colbeck and R. Renner},
  title   = {A system's wave function is uniquely 
             determined by its underlying physical 
             state},
  journal = {New Journal of Physics},
  volume  = {19},
  pages   = {013016},
  year    = {2017},
  doi     = {10.1088/1367-2630/aa515c}
}

@misc{friederichtyagi_PartI,
      title={Can Quantum Field Theory be Recovered from Time-Symmetric Stochastic Mechanics? Part I: Generalizing the Liouville Equation}, 
      author={Simon Friederich and Mritunjay Tyagi},
      year={2026},
      eprint={2603.20399},
      archivePrefix={arXiv},
      primaryClass={quant-ph},
      url={https://arxiv.org/abs/2603.20399}, 
}

\end{document}